# Modified Coupled-mode Theory for the Absorption in Plasmonic Lattices

*Joshua T. Y. Tse\*, Shunsuke Murai, and Katsuhisa Tanaka*

Department of Material Chemistry, Graduate School of Engineering, Kyoto University, Katsura, Nishikyo-ku, Kyoto 6158510, Japan

Email: tse@dipole7.kuic.kyoto-u.ac.jp

**Abstract**

Surface lattice resonance supported on plasmonic nanoparticle arrays enhances light-matter interactions for applications such as photoluminescence enhancement. The photoluminescence process is enhanced through confining light beyond the diffraction limit and inducing stronger light-matter interaction. In this work, the absorption mechanisms of plasmonic nanoparticle arrays embedded with photoluminescent absorbers are analyzed. A modified coupled-mode theory that describes the optical behavior of the surface lattice resonance was developed and verified by numerical simulations. Based on the analytical model, different components of the absorption contributed by the nanoparticles and the absorbers as well as the resonant properties of each of the components are identified. The origin of difference in resonant behavior with different materials is also discovered by exploring the nearfield characteristics of surface lattice resonance composed with a variety of materials.

# 1. Introduction

Plasmonic nanoparticles which support localized surface plasmon resonance (LSPR) were known to possess extraordinarily strong light localization and EM field enhancement near the vicinity of the nanoparticles, which enables strong light-matter interaction.[1-4] When the nanoparticles are arranged in a periodic structure, it gives rise to a variety of distinct phenomena such as negative refractive index, wavefront modulation and magnetic metamaterials.[5-9] In particular, our study focuses on exploring the plasmonic-photonic hybrid mode called surface lattice resonance (SLR).[10-16] While LSPR have a relatively low Q-factor, SLR greatly improves on this aspect and numerous studies have shown that SLR can be engineered to have Q-factor orders of magnitude higher.[15-18] Surface lattice resonance also retains the strong light confinement and field enhancement properties of LSPR, which have led to a wide range of applications such as SLR-based surface-enhanced Raman scattering (SERS),[19,20] collectively enhanced IR absorption (CEIRA) spectroscopy,[21,22] lasing,[23,24] circular dichroism (CD) spectroscopy,[25,26] etc.

In our previous work, we have established a modified coupled-mode theory (CMT) that describes the SLR supported on lossless nanoparticle arrays embedded in an absorptive index-matching layer.[27] In the framework of CMT, the EM field is decomposed into eigenmodes of the resonator and spatial characteristics such as the field enhancement pattern and the effective modal volume are incorporated into the parameters of the harmonic oscillators that represent each eigenmode. This grants CMT great versatility on describing a wide variety of structures such as core-shell nanoparticles and anisotropic materials. Based on the well-established lossless CMT, we introduced additional factors signatured by the non-resonant absorption of the absorptive index-matching layer to modify the dynamics of the SLR.[27,28] In short, the non-resonant absorption of the index-matching layer dampens the excitation and radiative decay of the SLR, as well as reducing the total intensity of the direct scattering. The modified CMT was shown to describe SLR supported on $TiO_2$ nanoparticle arrays embedded in a fluorescent index-matching layer very well with experimental and numerical evidence. The modified CMT also predicted the absorption of such systems accurately and provided an intuitive guide to efficient energy transfer from excitation beam to emitters in photoluminescence applications. However, this modified CMT is incompatible with plasmonic lattices in which the metallic nanoparticles exhibit intrinsic ohmic absorption. The extraordinarily strong field enhancement provided by plasmonic nanoparticles over dielectric nanoparticles is anticipated to further enhance the absorption of the fluorescent layer and improve the energy transfer efficiency.[27,29,30] In order to utilize this strong light-matter interaction provided by plasmonic nanoparticles, a comprehensive theory that describes the absorption mechanisms in such systems with two competing components is needed.

In this work, we propose an analytical model based on the coupled-mode theory to analyze the energy dissipation pattern when a lossy nanoparticle array is coupled with an absorptive index-matching layer. We then verify our CMT model with full-wave numerical simulation results based on the FDTD method. We further analyze the behavior of the non-resonant absorption and the absorptive decay mechanisms through the insights provided by our CMT model and with the help of the nearfield of the SLR obtained from FDTD simulations. Lastly, we discuss on the choice of materials capable of best enhancing photoluminescence in photoluminescence applications.

## 2. Modified Coupled-mode Theory for Plasmonic Nanoparticles

In our previous work, we have derived a modified CMT that describes the effect of an absorptive index-matching layer on a lossless nanoparticle array.[27] In our modified CMT, the mode amplitude $a$ of the SLR mode can be described by the following equation:[27,31-33] $\frac{d}{dt}a = \left(i\omega_0 - \frac{\Gamma_{tot}}{2}\right)a + \sqrt{\frac{\Gamma_{rad}}{2}}\alpha\langle\kappa^*|s_+\rangle$, and the outgoing wave $|s_-\rangle$ is described by $|s_-\rangle = \beta\mathbf{C}|s_+\rangle + a\sqrt{\frac{\Gamma_{rad}}{2}}\alpha|\kappa\rangle$. $\omega_0$ is the resonant frequency, $\Gamma_{tot}$ is the total decay rate and $\Gamma_{rad}$ is the radiative decay rate of the SLR, where the total decay rate is the sum of the radiative decay rate and the absorptive decay rate $\Gamma_{abs}$. $|s_+\rangle$ is the incident wave, $\mathbf{C}$ is the direct scattering matrix, and $|\kappa\rangle$ are the in-coupling constants. The modification terms $\alpha$ and $\beta$ are given by $\alpha = \sqrt[4]{1-A_0}$ and $\beta = \sqrt{1-A_0}$ where $A_0$ is the non-resonant absorptivity of the absorptive index-matching layer. We note that $\alpha$ and $\beta$ would return to 1 given a non-absorptive index-matching layer, which ensures consistency with the well-established lossless CMT. The TE/TM-polarized transmissivity and reflectivity were derived to be:

$$T_{TE/TM} = (1-A_0)\left|t_{0,TE/TM} + \frac{\Gamma_{rad}}{2}\frac{\kappa_{TE/TM}^2}{i(\omega-\omega_0)+\Gamma_{tot}/2}\right|^2 \text{ and} \quad (1)$$

$$R_{TE/TM} = (1-A_0)\left|r_{0,TE/TM} + \frac{\Gamma_{rad}}{2}\frac{\kappa_{TE/TM}^2}{i(\omega-\omega_0)+\Gamma_{tot}/2}\right|^2, \quad (2)$$

where $t_{0,TE/TM}$ and $r_{0,TE/TM}$ are the non-resonant transmissivity and reflectivity, and $\kappa_{TE/TM}$ are the coupling constants.

In order to examine the absorption of the dye and the nanoparticle independently, we assume the absorptive decay rate and the non-resonant absorptivity can be separated into components contributed by the dye and the nanoparticle respectively, denoted as $\Gamma_{abs} = \Gamma_{abs,dye} + \Gamma_{abs,NP}$ and $A_0 = A_{0,dye} + A_{0,NP}$. The absorptivity of the nanoparticle array by considering $A = 1 - \frac{\langle S_-|S_-\rangle}{\langle S_+|S_+\rangle}$ was derived as:

$$A = A_0 + (1 - A_0) \frac{2\Gamma_{rad}\Gamma_{abs}}{\Gamma_{tot}^2} \frac{|\langle \kappa^*|S_+\rangle|^2}{\langle S_+|S_+\rangle} \tag{3}$$

at the resonant wavelength of the SLR mode. We then use Equation 3 to predict the absorptivity contributed by the dye or the nanoparticle by substituting $\Gamma_{abs}$ and $A_0$ with the corresponding components. Since the sum of the absorptivity by the dye $A_{dye}$ and the nanoparticle $A_{NP}$ should return the total absorptivity, that is $A = A_{dye} + A_{NP}$, we find the components to be:

$$A_{dye} = A_{0,dye} + (1 - A_0) \frac{2\Gamma_{rad}\Gamma_{abs,dye}}{\Gamma_{tot}^2} \frac{|\langle \kappa^*|S_+\rangle|^2}{\langle S_+|S_+\rangle}, \tag{4}$$

$$A_{NP} = A_{0,NP} + (1 - A_0) \frac{2\Gamma_{rad}\Gamma_{abs,NP}}{\Gamma_{tot}^2} \frac{|\langle \kappa^*|S_+\rangle|^2}{\langle S_+|S_+\rangle}. \tag{5}$$

Note that when lossless nanoparticles are used, $A_{NP}$ goes to zero with $A_{0,NP} = 0$ and $\Gamma_{abs,NP} = 0$, and Equation 4 is identical to Equation 3, resembling our previous work. From Equation 4 and 5, we can see that as we increase the dye concentration, which consequently increases $A_{0,dye}$ and $\Gamma_{abs,dye}$, the non-resonant absorption term increases linearly with $A_{0,dye}$. However, the increase in SLR absorption is dampened by the $(1 - A_0)$ factor as well as the denominator $\Gamma_{tot}^2$, which also increases with $\Gamma_{abs,dye}$ since $\Gamma_{tot} = \Gamma_{rad} + \Gamma_{abs,dye} + \Gamma_{abs,NP}$. As a result, $A_{dye}$ is expected to follow a concave downwards pattern against increasing dye concentration. On the other hand, we expect $A_{0,NP}$ and $\Gamma_{abs,NP}$ to remain constant when only the dye concentration is changed. Since $A_{NP}$ also suffers from the dampening of the $(1 - A_0)$ factor and the denominator $\Gamma_{tot}^2$, $A_{NP}$ is expected to decrease monotonically as the dye concentration increases.

## 3. Absorption in Plasmonic Lattices

We verified our model by numerical simulations with nanoparticle arrays as illustrated in Figure 1a. The structures were based on our previous experimental works on the nanoparticle arrays with the emitter layers.[27] The Al cylindrical nanoparticles have height $H = 100$ nm and diameter $D = 90$ nm, and are placed in a square lattice with periodicity $P = 380$ nm. (This structure will be labelled as Al-D90H100-Sq.) The refractive index of the nanoparticles was obtained from Ref [34]. The lattice was placed on a substrate with refractive index $n = 1.46$ and embedded into an index-matching layer with thickness $t = 280$ nm and complex refractive index $n = 1.46 + i\kappa$. A semi-infinite layer of vacuum is placed above the index-matching layer. The extinction coefficient $\kappa$ is introduced to simulate the absorption introduced by absorbers distributed inside the index-matching layer at various concentrations. The $x$- and $y$-boundaries of the unit cell were terminated by periodic boundary conditions to simulate an infinite array, while perfectly-matching layers were used to terminate the z-boundaries to simulate semi-infinite extension of the substrate and superstrate. A broadband plane wave source was used to excite SLR under normal incident from the substrate side. Power monitors were placed above and below the structure to detect the transmissivity $T$ and reflectivity $R$ of the lattice while the absorptivity $A$ is calculated by $A = 1 - T - R$. The Poynting vector at all mesh points in the nanoparticle were also recorded and the divergence of the Poynting vector in the nanoparticle was integrated to obtain the absorptivity by the nanoparticle $A_{NP}$. The absorptivity by the dye was then calculated by $A_{dye} = A - A_{NP}$.

Figure 1b–f shows the simulated transmissivity, reflectivity and absorptivity spectra at normal incident for $\kappa$ = 0, 0.0063, 0.0126, 0.0189, 0.0252. The best fit with the CMT were also plotted in Figure 1 as the black dashed lines. The total, radiative and absorptive decay rates obtained from the best fits are plotted against $\kappa$ in Figure 2a and the $A_0$ is also plotted against $\kappa$ in Figure 2b. From the fitted parameters, we can see that while the radiative decay rate $\Gamma_{rad}$ is almost constant against $\kappa$, the absorptive decay rate $\Gamma_{abs}$ and the $A_0$ both linearly increase with $\kappa$. This results in a linear increase of the total decay rate $\Gamma_{tot}$ with $\kappa$. In Figure 2a, we observe an intersection between $\Gamma_{rad}$ and $\Gamma_{abs}$ at $\kappa$ = 0.0068, which indicates that the term $\frac{2\Gamma_{rad}\Gamma_{abs}}{\Gamma_{tot}^2}$ is maximized. At $\kappa$ = 0, the $\Gamma_{abs}$ and $A_0$ are both non-zero, which indicates that the nanoparticles contributed to the absorption as no dye is present in that configuration. We distinguish between the absorption by the dye and by the nanoparticles by fixing the absorptive decay rate and the non-resonant absorptivity contribution of the nanoparticle to the values at $\kappa$ = 0, that is $\Gamma_{abs,NP} = \Gamma_{abs}(\kappa = 0)$ and $A_{0,NP} = A_0(\kappa = 0)$. As $\kappa$ increases, the change in $\Gamma_{abs}$ and $A_0$ from $\kappa$ = 0 is attributed solely to the dye, as illustrated in Figure 2a–b. We proceed to use the linear regression of the obtained parameters to predict $A$, $A_{dye}$ and $A_{NP}$ with Equation 3 – 5. The simulated $A$, $A_{dye}$ and $A_{NP}$ at the resonant wavelength of the SLR are plotted against $\kappa$ in Figure 2c, where the CMT model was shown to predict each absorption component accurately. As shown in Figure 2c, while the $A_{dye}$ follows a monotonically increasing trend, the $A_{NP}$ shows an opposite trend and decreases quickly as the dye is introduced to the system. As a result, the absorptivity follows a hump shape with the peak at 0.56 at $\kappa$ = 0.0093.

To further show the versatility of our CMT model, we performed additional simulations on a variety of system configurations. As summarized in Table 1, nanoparticles with different diameter $D$ = 80, 130 nm (labelled as Al-D80H100-Sq, Al-D130H100-Sq) and height $H$ = 50 nm (labelled as Al-D90H50-Sq), a hexagonal nanoparticle array ($P$ = 440 nm, labelled as Al-D90H100-Hex) which has identical SLR positions to $P$ = 380 nm square lattices at normal incidence, and a Ag nanoparticle array ($D$ = 90 nm, labelled as Ag-D90H100-Sq). All other simulation parameters are identical to the parameters used in the primary configuration (Al-D90H100-Sq). The fitted $\Gamma_{tot}$, $\Gamma_{rad}$ and $\Gamma_{abs}$ are plotted as a function of $\kappa$ in Figure 3 and the fitted $A_0$ are plotted in Figure 4. As shown in Figure 3, the $\Gamma_{rad}$ only show minimal changes as $\kappa$ increases, while the $\Gamma_{tot}$ and $\Gamma_{abs}$ increase linearly with $\kappa$, similar to Al-D90H100-Sq. In particular, we can see that $\Gamma_{rad}$ and $\Gamma_{abs}$ intersect at relatively small $\kappa$ for Al-D80H100-Sq, Al-D90H50-Sq and Al-D90H100-Hex (0.0017, 0.0004 and 0.006 respectively), while the intersection of Al-D130H100-Sq and Ag-D90H100-Sq occurs at significantly larger $\kappa$, at 0.0506 and 0.0245 respectively.

The parameters are then used to predict the $A$ of each configuration, as plotted in Figure 5. The simulated absorptivities are also plotted in Figure 5 to show the model accurately predict the total absorptivity and the absorptivity by each component. The configurations Al-D80H100-Sq, Al-D90H50-Sq and Al-D90H100-Hex displayed similar behavior to Al-D90H100-Sq. While the $A$ shows a maximum, or so-called critical coupling,[27] at a relatively small $\kappa$, the $A_{dye}$ only exhibited a monotonic increase with $\kappa$. On the other hand, Al-D130H100-Sq and Ag-D90H100-Sq showed no peak value in both $A$ and $A_{dye}$. This difference in behaviour can be explained by the value of $\kappa$ at which $\Gamma_{rad}$ and $\Gamma_{abs}$ intersect, which indicates where the SLR absorption is maximized. Since the $A_0$ term and the $(1 - A_0)$ factor in Equation 3 effectively reduces the portion of the absorption that depends on the contribution of SLR, the subsequent roll-off of the absorption by SLR after the intersection is counteracted by the increase in $A_0$. As a result, the $A$ did not show any peak value but a monotonic increase when the intersection occurs at a large $\kappa$. Therefore, we conclude that our model is adequate in describing the decay mechanism and absorption of this type of nanoparticle arrays with an absorptive layer.

## 4. Discussion

We can learn more about the decay mechanisms of these SLR by comparing the trends of various parameters of different structures. The $A_0$ for each structure are plotted together in Figure 6a, and $A_{0,NP}$ and the slope of $A_{0,dye}$ of each structure are plotted in Figure 6c for comparison. (We also include the TiO$_2$ nanoparticle array from Ref [27] in the comparison, labelled TiO$_2$-D130H100-SQ.) As shown in Figure 6a, the $A_0$ of each structure only differ by a vertical displacement and the slope of its increase along $\kappa$ is independent of the array structure. This can be analyzed by considering $A_{0,dye}$ and $A_{0,NP}$ separately. $A_{0,dye}$ is primarily contributed by the absorption of the dye layer when there are no nanoparticles embedded inside and the influence of the nanoparticle is relatively small. Maxwell's equations predict the empty layer should give a slope of 6.34, from which the fitted slopes of $A_{0,dye}$ are smaller than. The difference can be attributed to the area occupied by the nanoparticle, which would reduce the $A_{0,dye}$ from the empty layer approximation. In particular, we can see that the structures with the largest nanoparticles (Al-D130H100-Sq and TiO$_2$-D130H100-Sq) show the most reduction in $A_{0,dye}$ while Al-D80H100-Sq and Al-D90H100-Hex with lower packing density show the least reduction. On the other hand, $A_{0,NP}$ captures the non-resonant effect of the nanoparticle, which offsets the $A_0$ due to the difference in packing density and absorption cross section of the nanoparticle. The effect of the packing density can be illustrated by the difference in $A_{0,NP}$ between Al-D90H100-Sq and Al-D90H100-Hex. Both share the same nanoparticle but Al-D90H100-Hex has a packing density 13.9% smaller than Al-D90H100-Sq, and the $A_{0,NP}$ is also 14% smaller. We further generalize by comparing $A_{0,NP}$ with the absorption cross section $\sigma_{abs}$ of individual nanoparticles normalized to the unit cell area, which we find $\sigma_{abs}/area$ follows a similar trend as $A_{0,NP}$. This suggests that $A_{0,NP}$ originates from the absorption of individual nanoparticles before forming SLR, i.e. the absorption from LSPR.

We also compared the trend of the $\Gamma_{abs}$ to explore the absorption mediated by the SLR. The $\Gamma_{abs}$ of each structure are plotted in Figure 6b,[27] which shows that $\Gamma_{abs}$ strongly depends on what nanoparticle is placed in the lattice. The $\Gamma_{abs,NP}$ shows a clear distinction between different materials, where TiO$_2$-D130H100-Sq has a $\Gamma_{abs,NP}$ close to zero, and Ag-D90H100-Sq shows a significantly larger $\Gamma_{abs,NP}$ than the Al nanoparticle arrays. On the other hand, the $\Gamma_{abs,dye}$ points towards a more intricate trend. While Al-D130H100-Sq show the strongest $\Gamma_{abs,dye}$, Ag-D90H100-Sq and other Al nanoparticle arrays have a similar $\Gamma_{abs,dye}$. The TiO$_2$-D130H100-Sq also shows a distinctly smaller $\Gamma_{abs,dye}$. Since $\Gamma_{abs}$ relates the power absorbed $P_{abs}$ to the mode amplitude of the SLR by $P_{abs} = \Gamma_{abs}|a|^2$, where $|a|^2$ is normalized to the total optical energy stored in the mode, we can compare the $|E_{SLR}|^2$ inside the absorptive dye layer to the total $|E_{SLR}|^2$ of the SLR mode to estimate the trend of $\Gamma_{abs,dye}$, i.e. $\frac{\Gamma_{abs,dye}}{\kappa} \propto \frac{\int_{dye}|E_{SLR}(r)|^2 d^3r}{\int_{all}\varepsilon'(r)|E_{SLR}(r)|^2 d^3r}$, where $\varepsilon'(r)$ is the real part of the relative permittivity at position $r$.[35,36] The nearfield patterns from the numerical simulations at $\kappa = 0$ are plotted in Figure 7, in which we can see that the nearfield of TiO$_2$-D130H100-Sq shows a wide spread across $z$ from $z = -350$ nm to 250 nm. In contrast, the nearfields of the Al nanoparticles mainly span from $z = -30$ nm to 130 nm, while Ag-D90H100-Sq constrains the nearfield even more to only around $z = -10$ nm to 120 nm. The fields are also confined close to the top and bottom edges of the Al and Ag nanoparticles, which is typical for plasmonic nanostructures. We can also see in Figure 7 where while a significant portion of the field is trapped within the TiO$_2$ nanoparticle, the field did not penetrate deep into the Al and Ag nanoparticles. Therefore, the overall trends suggest that the TiO$_2$ is worst at concentrating the EM field into the index-matching layer, with a significant portion of the field remaining in the substrate as well as in the nanoparticles. In contrast, the Ag enhances the field near the nanoparticles more, thus resulting in a more localized nearfield in the proximity of the nanoparticle. We then integrate $|E_x(\mathbf{r})|^2 + |E_z(\mathbf{r})|^2$ over the volume of the index-matching layer and the whole unit cell respectively to estimate the portion of optical energy of the SLR confined inside the index-matching layer. ($E_y$ is omitted due to interference from the incoming and outgoing waves.) The ratio of the integrated fields $\frac{\int_{dye} |E_x(\mathbf{r})|^2 + |E_z(\mathbf{r})|^2 d^3r}{\int_{all} \varepsilon'(\mathbf{r})(|E_x(\mathbf{r})|^2 + |E_z(\mathbf{r})|^2) d^3r}$ and the slope of $\Gamma_{abs,dye}$ are plotted with in Figure 6d. Overall, we can see that the estimations from the integrated E-fields show good consistency over different materials and structures with the fitted $\Gamma_{abs,dye}$. In particular, when comparing Al-D130H100-Sq and TiO$_2$-D130H100-Sq, we found that Al nanoparticles almost doubled the $\Gamma_{abs,dye}$ compared to TiO$_2$. Switching to Ag nanoparticles would also slightly decrease $\Gamma_{abs,dye}$ from Al nanoparticles, as seen in the comparison between Al-D90H100-Sq and Ag-D90H100-Sq.

Although the plasmonic nanoparticles can induce much stronger $\Gamma_{abs,dye}$ when compared to TiO$_2$, the resultant fluorescent absorption efficiency was worse than that of the TiO$_2$ nanoparticle array in Ref [27]. This is due to the existence of $\Gamma_{abs,NP}$ to compete with $\Gamma_{abs,dye}$ for the optical energy stored in the SLR mode. Since $\Gamma_{tot} = \Gamma_{rad} + \Gamma_{abs,dye} + \Gamma_{abs,NP}$, the optical energy stored in the SLR mode can decay through either of the three pathways, namely radiative decay, absorption by dye and absorption by nanoparticle. While the radiative decay can be controlled by engineering the incident light, the dissipation by $\Gamma_{abs,NP}$ cannot be easily eliminated on plasmonic nanoparticles. The only way to reduce the contribution of $\Gamma_{abs,NP}$ is to increase $\Gamma_{abs,dye}$ such that $\Gamma_{abs,dye}$ is dominant in $\Gamma_{abs}$. However, this would also require a large $\Gamma_{rad}$ to effectively in-couple such a large power and it is not straightforward to strongly increase $\Gamma_{rad}$ without significantly affecting other parameters, such as $\Gamma_{abs,NP}$. Also, the required $\kappa$ may not be suitable for photoluminescence applications as the larger $A_{0,dye}$ would impede the energy transfer to SLR, and may lead to other undesirable effects such as concentration quenching in dyes. Al is also known to suffer from rapid oxidation when in contact with ambient air and forms a thin protective Al$_2$O$_3$ layer on its surface. Since amorphous Al$_2$O$_3$ is a non-absorptive dielectric material, it has limited effect on the modal properties of SLR. However, since the Al$_2$O$_3$ still displaces the dye from the surface of the Al, which is where the enhancement is the strongest, the fluorescent enhancement provided by the plasmonic lattice would be reduced relative to the unoxidized case.

Interestingly, these field patterns also give us some insight into what type of materials are more suitable in applications for emission enhancement, which requires different qualities than absorption enhancement. Since the fractional radiative local density of states (LDOS) is directly related to the average nearfield enhancement, we can connect the nearfield enhancement patterns with spontaneous emission characteristics through the Lorentz reciprocity theorem: a stronger E-field at the emitters when incident with $\boldsymbol{k}$ is equivalent to a stronger emission enhancement at emission wavevector $-\boldsymbol{k}$.[37-39] Therefore, instead of having lower absorption, which is beneficial for fluorescent absorption enhancement, the nanoparticle arrays should better confine the E-field of the SLR mode into the emitter layer in order to maximize the emission at that specific direction, which is same as the conditions of best enhancing $\Gamma_{abs,dye}$. Therefore, we can infer that the TiO$_2$ nanoparticle array is less effective on emission enhancement applications when compared to the plasmonic counterparts. The relative position between the emitters and the nanoparticles are also crucial. Since the plasmonic nanoparticles focus the enhancement into the vicinity of the top and bottom edge of the nanoparticles, the plasmonic lattice is more sensitive in the position of the emitters and is most effective when the emitters are close to the nanoparticles. On the other hand, the dielectric lattice provides enhancement over a larger volume and enhances evenly over the whole volume of the emitter layer.

## 5. Conclusion

In conclusion, we derived the modified CMT that describes the energy dissipation pattern when a plasmonic lattice is coupled with an absorptive index-matching layer. We identified the contributions of the dye and nanoparticle in the absorption decay rates and the non-resonant absorption respectively and utilized this distinction to predict the energy absorption by each component. We then verified our modified CMT with FDTD simulation results to demonstrate the versatility of our model in describing a variety of configurations. Finally, we analyzed the behavior of the non-resonant absorption and the resonant absorptive decay through the insights provided by the modified CMT. We also discussed the effect of the nearfield of the SLR on the absorption efficiency improvements and implications on the choice of materials for photoluminescence enhancement through relating LDOS with the conditions behind the absorptive decay rate by dye.

**Acknowledgements**

The authors acknowledge financial support from Kakenhi (22H01776, 22K18884, 21H04619), MEXT, Japan.

**Figure 1.** (a) The illustration of the nanoparticle array used in FDTD simulations. The nanoparticles are cylindrical with height $H$ and diameter $D$, and are placed in a square lattice with periodicity $P$. The nanoparticle array is covered by an index-matching layer of thickness $t$. (b) – (f) The FDTD simulated transmissivity $T$, reflectivity $R$ and absorptivity $A$ spectra at normal incident for $\kappa$ = (b) 0, (c) 0.0063, (d) 0.0126, (e) 0.0189 and (f) 0.0252. The best fits with the CMT are plotted as the dashed lines.

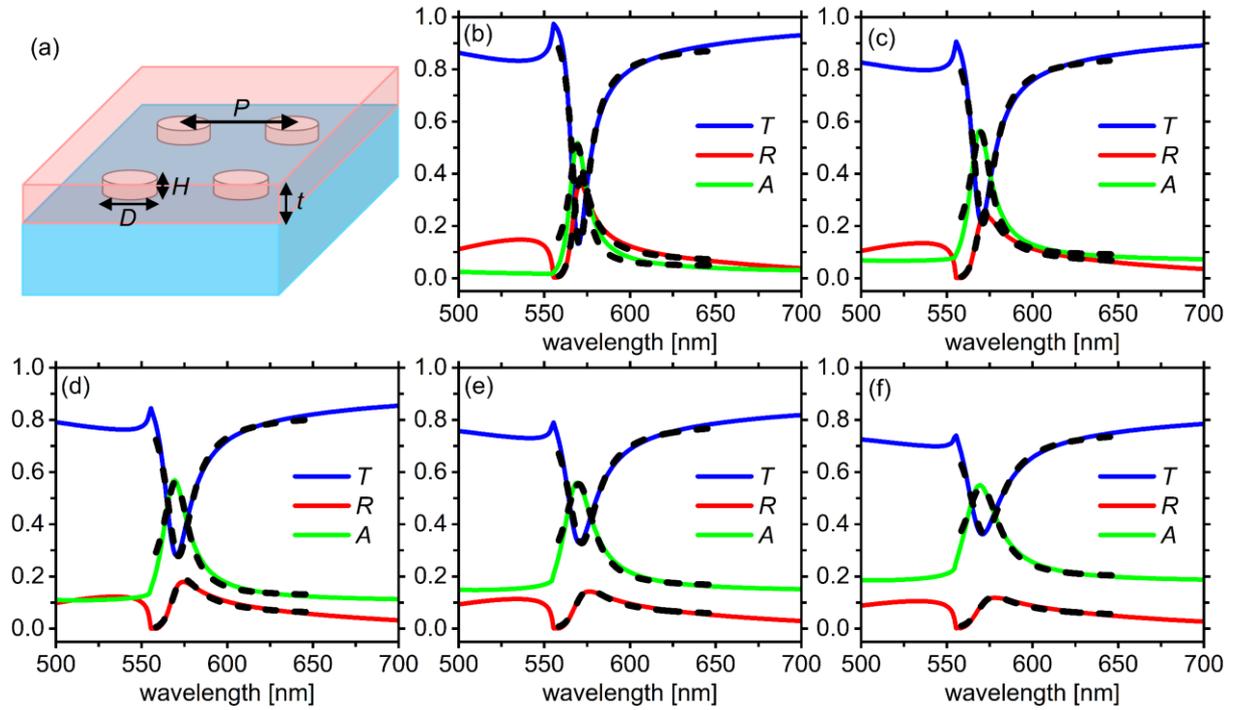

**Figure 2.** The (a) $\Gamma_{tot}$ (black squares), $\Gamma_{rad}$ (red circles), $\Gamma_{abs}$ (blue triangles) and (b) $A_0$ of Al-D90H100-Sq are plotted as a function of $\kappa$. The trends of the parameters are fitted by the straight lines as shown. The grey dashed lines and arrows illustrate the attribution of $\Gamma_{abs,NP}$, $\Gamma_{abs,dye}$, $A_{0,NP}$ and $A_{0,dye}$. (c) The simulated $A$ (black squares), $A_{dye}$ (red circles) and $A_{NP}$ (blue triangles) at the resonant wavelength of the SLR are plotted as a function of $\kappa$. The solid lines are the CMT predicted absorptivity from the fittings.

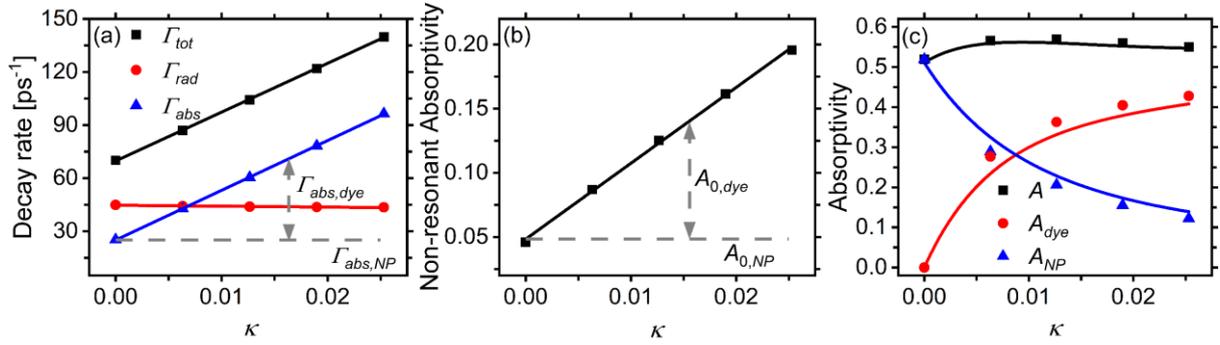

**Table 1.** The geometrical parameters used in the simulations of different nanoparticle arrays.

| $D$ [nm] | $H$ [nm] | $P$ [nm] | Lattice type | Material | Label |
|---|---|---|---|---|---|
| 90 | 100 | 380 | square | Al | Al-D90H100-Sq |
| 80 | 100 | 380 | square | Al | Al-D80H100-Sq |
| 130 | 100 | 380 | square | Al | Al-D130H100-Sq |
| 90 | 50 | 380 | square | Al | Al-D90H50-Sq |
| 90 | 100 | 440 | hexagonal | Al | Al-D90H100-Hex |
| 90 | 100 | 380 | square | Ag | Ag-D90H100-Sq |
| 130 | 100 | 380 | square | $TiO_2$ | $TiO_2$-D130H100-Sq |

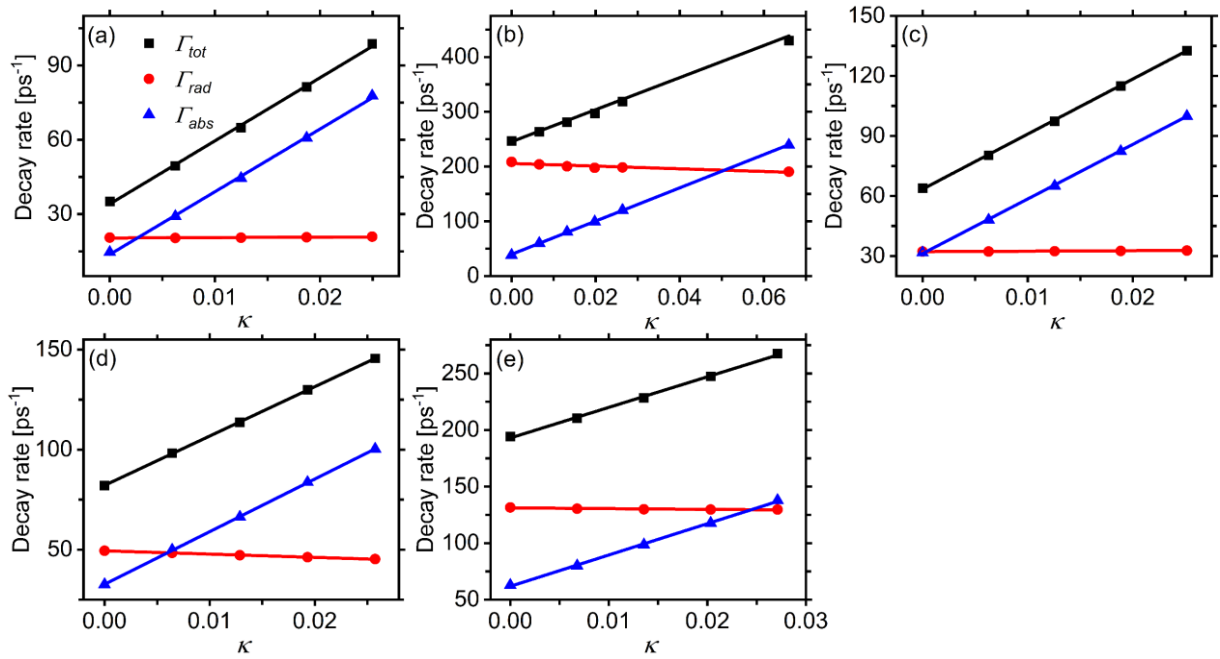

**Figure 3.** The fitted $\Gamma_{tot}, \Gamma_{rad}, \Gamma_{abs}$ of configurations (a) Al-D80H100-Sq, (b) Al-D130H100-Sq, (c) Al-D90H50-Sq, (d) Al-D90H100-Hex, and (e) Ag-D90H100-Sq are plotted as a function of $\kappa$. The trends of the parameters are fitted by the straight lines as shown.

**Figure 4.** The fitted $A_0$ of configurations (a) Al-D80H100-Sq, (b) Al-D130H100-Sq, (c) Al-D90H50-Sq, (d) Al-D90H100-Hex, and (e) Ag-D90H100-Sq are plotted as a function of $\kappa$. The trends of the parameters are fitted by the straight lines as shown.

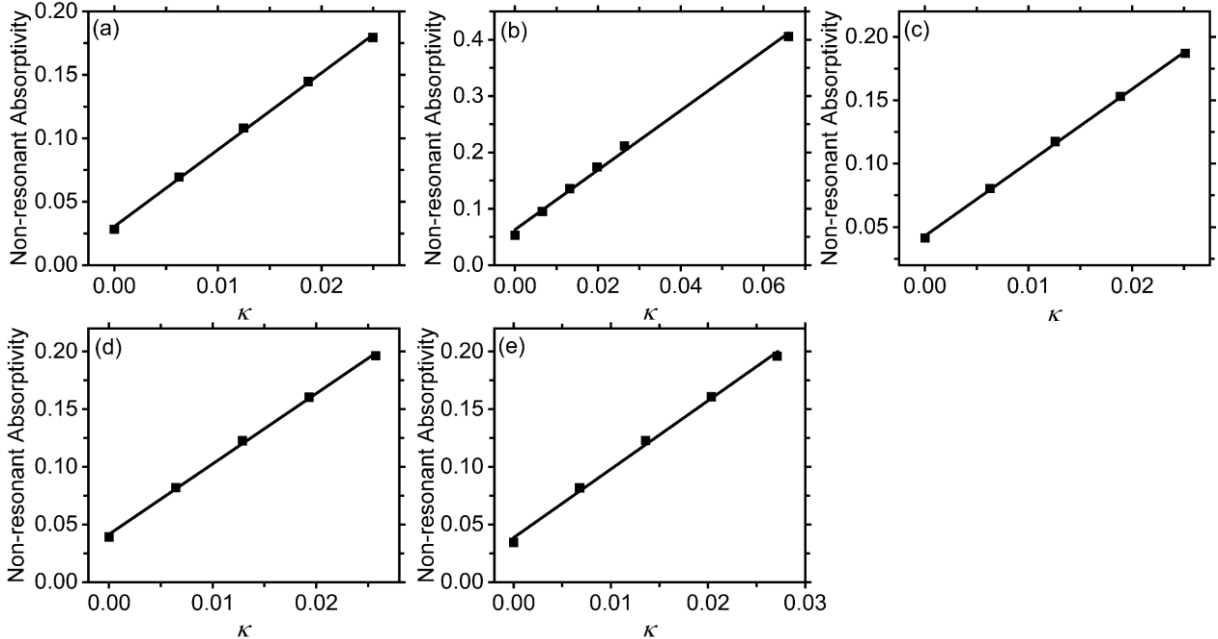

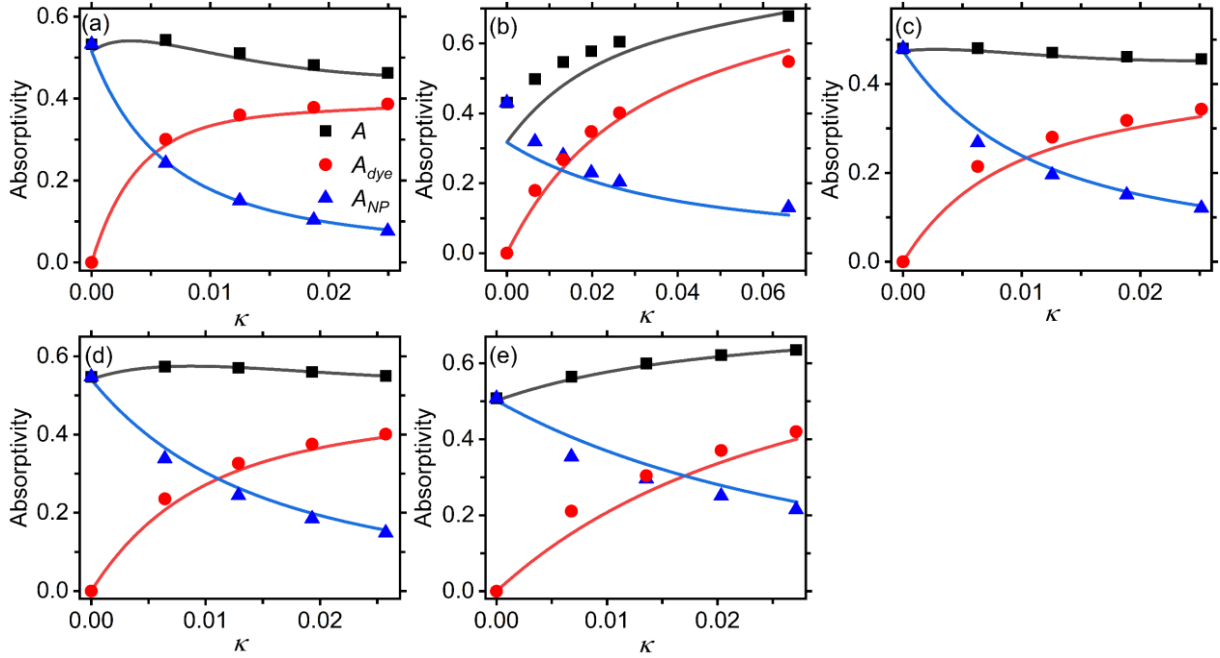

**Figure 5.** The CMT predicted and the FDTD simulated $A$, $A_{dye}$ and $A_{NP}$ at the resonant wavelength of the SLR of configurations (a) Al-D80H100-Sq, (b) Al-D130H100-Sq, (c) Al-D90H50-Sq, (d) Al-D90H100-Hex, and (e) Ag-D90H100-Sq are plotted as a function of $\kappa$. The solid lines show the CMT predictions while the symbols show the FDTD simulated results.

**Figure 6.** The (a) $A_0$ and (b) $\Gamma_{abs}$ of each structure are plotted as a function of $\kappa$. (c) The $A_{0,NP}$ (black squares), the $\sigma_{abs}/area$ (blue triangles) and the slope of $A_{0,dye}$ (red circles) of each structure are plotted. The empty layer approximation for the slope of $A_{0,dye}$ is indicated by the red dashed line. (d) The slope of $\Gamma_{abs,dye}$ (black squares) of each structure are plotted with the ratio of the integrated nearfields $\dfrac{\int_{dye}|E_x(r)|^2+|E_z(r)|^2 d^3r}{\int_{all}\varepsilon\prime(r)(|E_x(r)|^2+|E_z(r)|^2)\,d^3r}$ (red circles). The spectrally fitted $\Gamma_{abs,dye}$ shows good agreement with the estimation from integrated nearfields.

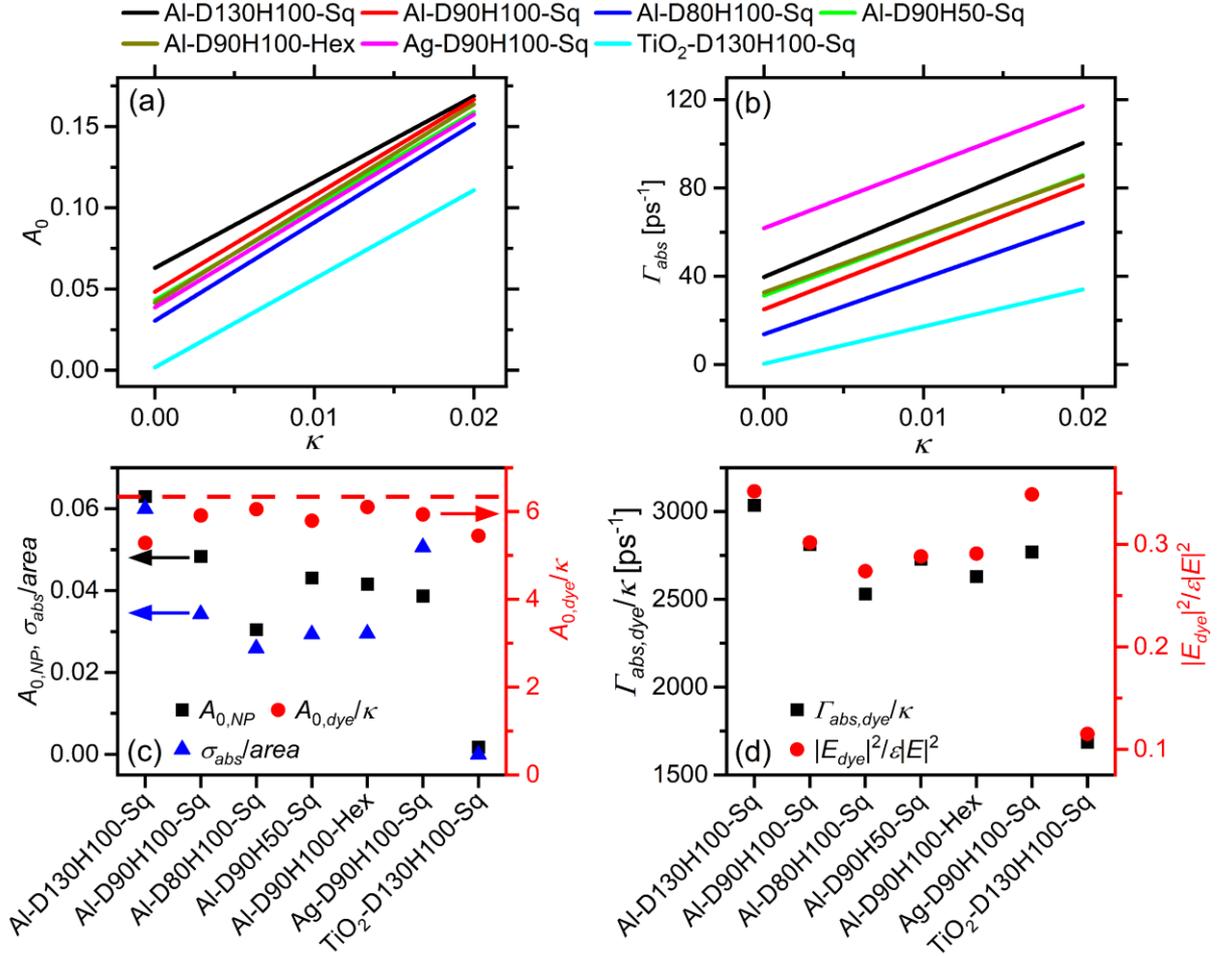

**Figure 7.** The nearfield pattern of configurations (a) Al-D90H100-Sq, (b) Al-D80H100-Sq, (c) Al-D130H100-Sq, (d) Al-D90H50-Sq, (e) Al-D90H100-Hex, (f) Ag-D90H100-Sq, and (g) TiO$_2$-D130H100-Sq are plotted at the respective resonant wavelength of the SLR. The E-fields normalized to the incident field strength are plotted as a function of $y$ and $z$ along the center of the nanoparticle. The black lines indicate the edge of the nanoparticles.

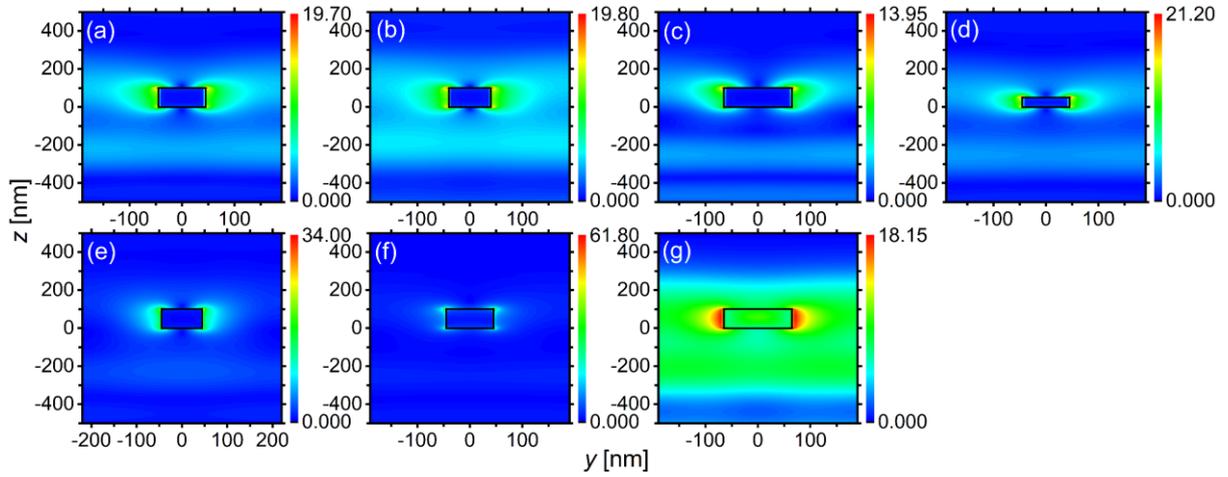